\documentclass[conference]{IEEEtran}
\usepackage[draft]{hyperref}
\usepackage[noadjust]{cite}
\usepackage[pdftex]{graphicx}
\usepackage[cmex10]{amsmath}
\usepackage{algorithmic}
\usepackage{array}
\ifCLASSOPTIONcompsoc
  \usepackage[caption=false,font=normalsize,labelfont=sf,textfont=sf]{subfig}
\else
  \usepackage[caption=false,font=footnotesize]{subfig}
\fi
\usepackage{dblfloatfix}
\usepackage{url}
\usepackage[listings,skins]{tcolorbox}
\tcbuselibrary{fitting}
\usepackage{pdfcomment}
\usepackage{sidecap}
\usepackage{lineno}
\newtcolorbox[auto counter, number within=subsection]{pabox}[2][]{%
  title={\footnotesize Example~\thetcbcounter: #2}, #1}

\begin{document}
\title{A Community Contribution Framework for Sharing Materials Data with
  Materials Project 
  }
\author{%
  \IEEEauthorblockN{\textit{Patrick Huck}, Anubhav Jain, Dan Gunter, Donald Winston, Kristin Persson}
  \IEEEauthorblockA{%
    Energy Storage \& Distributed Resources Division\\%
    Lawrence Berkeley National Laboratory\\%
    Berkeley, California 94720\\%
    \{phuck, ajain, dkgunter, dwinston, kapersson\}@lbl.gov}%
}
\maketitle

\begin{abstract}
As scientific discovery becomes increasingly data-driven, software platforms
are needed to efficiently organize and disseminate data from disparate sources.
This is certainly the case in the field of materials science. For example,
Materials Project has generated computational data on over 60,000 chemical
compounds and has made that data available through a web portal and REST
interface. However, such portals must seek to incorporate community submissions
to expand the scope of scientific data sharing. In this paper, we describe
\texttt{MPContribs}, a computing/software infrastructure to integrate and
organize contributions of simulated or measured materials data from users. Our
solution supports complex submissions and provides interfaces that allow
contributors to share analyses and graphs. A RESTful API exposes mechanisms for
book-keeping, retrieval and aggregation of submitted entries, as well as
persistent URIs or DOIs that can be used to reference the data in publications.
Our approach isolates contributed data from a host project's quality-controlled
core data and yet enables analyses across the entire dataset, programmatically
or through customized web apps. We expect the developed framework to enhance
collaborative determination of material properties and to maximize the impact
of each contributor's dataset. In the long-term, \texttt{MPContribs} seeks to
make Materials Project an institutional, and thus community-wide, memory for
computational and experimental materials science.
\end{abstract}

\IEEEpeerreviewmaketitle

\section{Introduction}

As a key player in the U.S. Materials Genome Initiative~\cite{WHMGI2014},
Materials Project (MP)~\cite{Jain2013} uses High-Performance Computing
(HPC) to determine structural, thermodynamic, electronic, and mechanical
properties of over 60,000 inorganic compounds by means of high-throughput
ab-initio calculations. The calculation results and analysis tools are
disseminated to the public via modern web and application interfaces. These
results and tools serve to accelerate the discovery, design and creation of
next-generation materials for applications such as batteries, photovoltaics,
and semiconductors.

However, the materials science research community has a
continually increasing supply of experimental and theoretical material
properties that are either not yet calculated by MP or outside the scope of
MP's data generation efforts. Therefore, with a growing user base of over
10,000 registered users, it becomes increasingly important for MP and
similar scientific platforms to also enable community-driven submissions, which
would extend the scope of the possible applications and improve the
integrity/quality of the provided datasets, hence enhancing its value to the
user community.

The main contribution of this paper is a methodology for integrating and
organizing an existing materials dataset with community contributions. We
choose to approach this problem statement within the context of the Materials
Project for two reasons: first, the solution discussed here represents an
abstract solution in response to a specific problem for this initiative, rather
than a specific solution in search of abstract problems. Second, we emphasize
the ``live'' nature of our solution, which serves a large and existing user
community and maintains an existing database as part of the initiative.

Our solution includes the following novel
elements: (i) A syntax, \texttt{MPFile}, that allows users to easily describe
computed or experimental data; (ii) A mechanism for hierarchically correlating
user data with material(s) in the core project database; and (iii), A REST API
and UI framework for disseminating results to the community via customizable
interactive graphs.

The paper is organized as follows: Section~\ref{subsec_intro_relwork} reviews
related work. Sections~\ref{subsec_intro_overview} through \ref{sec_webacc}
describe the major components of our framework. Sections~\ref{sec_sum} and~\ref{sec_out}
conclude the paper with a summary and outlook on the future.

\begin{figure*}
\centering
\includegraphics[width=\textwidth]{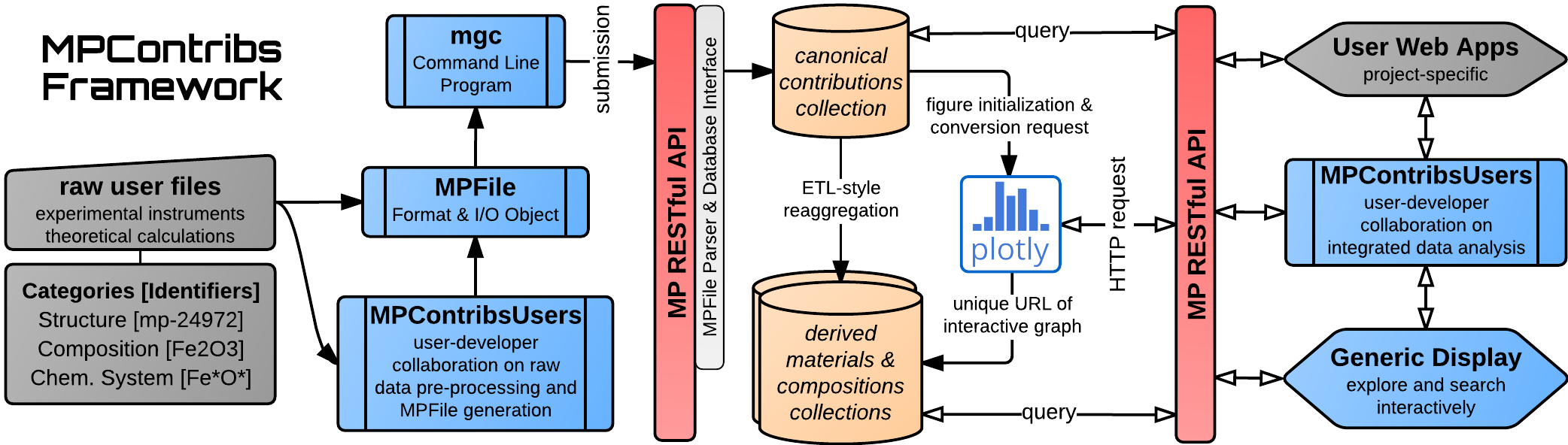}
\caption{Processing workflow of \texttt{MPContribs} framework. The
  pre-submission stages, with category identification and \texttt{MPFile}
  generation, are followed by internal parser and builder phases and the
  dissemination to the interactive portal. Contribution submission as well as
  queries to the databases (orange) and Plotly requests operate through the
  MP's REST API (red). Blue boxes and the processing behind the REST API ``wall''
  are provided by the \texttt{MPContribs} framework. Grey boxes denote tasks and
  tools under the user's responsibility. For more details, see text.}
\label{fig_flow_chart}
\end{figure*}

\section{Related Work}\label{subsec_intro_relwork}

There are a number of community repositories in Materials Science and related
disciplines. Traditionally, these repositories have been limited in scope to
crystal structures of measured compounds (acting analogously to the Protein
Data Bank~\cite{Berman2000} for biology), and include efforts such as the
Inorganic Crystal Structure Database (ICSD)~\cite{Belsky2002} 
and Pauling File~\cite{pauling}. More recent efforts, such as Materials
Project~\cite{Jain2013}, AFLOWlib~\cite{aflowlib}, and
others~\cite{cccbdb,comp-es,ortiz2009data,cepdb,oqmd}, typically include
additional property data (generally computed) but do not currently have a
contribution framework.  Some repositories, such as NoMaD~\cite{nomad},
iochem-bd~\cite{moreno2015managing}, and the Materials Data Facility
(MDF)~\cite{mdf}, provide a shared repository of data, with potentially large
experimental and structural data items, but many of these are in the early
stages of development and without integrated data analysis capabilities over a
core dataset. Our approach, by contrast, focuses on relatively small data
contributions \textit{that can be joined with the carefully curated core
  dataset} and extended with an integrated set of data analysis tools to
provide a materials design environment.

A number of projects in other sciences, particularly life sciences, share the
goals of community-enabling data contributions and analysis in the context of
reference datasets. The iPlant Collaborative~\cite{iplant} combines a ``Data
Store'' with a ``Discovery Environment'' with the goal of connecting public and
private datasets with analysis capabilities; the analysis capabilities include
workflows running on remote clusters and private Virtual Machines. The
DOE's Systems Biology Knowledgebase (KBase)~\cite{kbase}
provides a unified and interactive analysis environment built on the IPython
notebook~\cite{perez2007ipython} and a back-end store that can store users'
analyses together with both their data and standard reference microbial data.
The breadth of either of these projects exceeds the near-term goals of the
contribution framework described here, but point to the importance of this idea
across scientific communities.

Efforts such as ChemSpider~\cite{chemspider} emphasize aggregation of data from
many individual sources rather than extension of ``core'' reference data.
However, the success of such efforts depends on widely-used standards for
identifiers~\cite{Heller2013} and does not integrate community contributions as
added layers on top of a centrally generated and curated dataset. In contrast,
our approach enhances the value of both the canonical data (by adding community
contributions) and the community contributions (by adding the context of the
canonical data). Combining these datasets requires the development of new
methods. In the case of materials data/identifiers, the non-trivial problem
arises of how to connect the ``material'' as defined by the user, e.g. through
experimental characterization, with a ``material'' as defined by a
computational dataset.  Our classification of several types of contribution
(see Sec. \ref{subsec_preproc_matdef}) is a novel approach to ``divide and
conquer'' this type of problem.
  
In conclusion, most repositories in materials science that focus solely on
community contributions face difficulties in (i) building a user audience, (ii)
developing integrated search across datasets, and (iii) providing processed and
interpretable data (e.g., interactive plots) rather than raw data dumps. Our
integration of existing and novel datasets is a first attempt at addressing
such issues. Thus, our framework occupies a unique and important niche in the
materials data sharing space.

\section{Framework Overview \& Design}\label{subsec_intro_overview}

An overview of our framework \texttt{MPContribs}~\cite{mpcontribs} is shown in
Figure~\ref{fig_flow_chart}. From left to right, we see the processing workflow
for incorporating a user contribution. First, the raw user output is identified
and gathered. Examples of such data include:

\begin{itemize}
\item X-ray Absorption Spectroscopy (XAS) at light sources,
\item diffusivities computed for various temperatures \& solutes,
\item band gaps calculated under alternative conditions,
\item adsorption properties of nano-porous materials, and
\item electronic structure calculations for photovoltaics.
\end{itemize}

Next, the raw data is converted into a concise summary using the
\texttt{MPFile} syntax (Sec.~\ref{sec_preproc}). The result of this phase is
submitted, by the user, via the dedicated command line program \texttt{mgc}
(Sec.~\ref{subsec_mgc}) or manually through the web portal, both of which
operate through the RESTful MP API (Sec.~\ref{subsec_webacc_rest}). The
received data is parsed (Sec.~\ref{sec_parsbld}), and the resulting objects are
correlated with existing materials data via suitable identifiers. The results
are inserted into the core database. Finally, the data is disseminated through
the API---and a graphical UI (Sec.~\ref{subsec_webacc_frontend}) with
interactive graphs, tables and tree structures---for immediate exploration and
analysis.



\section{Submission Processing}\label{sec_preproc}

It is crucial to establish a common language for data interchange across
contributors from different research disciplines and to encourage managable
contribution sizes. During the pre-submission processing, the user's existing
(raw) materials data information is thus converted into a custom but general
input file format. This section describes in detail the format and processing
steps provided by \texttt{MPContribs} for this task.

\subsection{Material Definition \& Contribution Scope}\label{subsec_preproc_matdef}

A key challenge for the framework is to link together separate datasets that
pertain to the same material; this requires a universal comparator for the term
\textit{material}. MP defines a material as a group of structures that
share the same symmetry group as well as comparable lattice parameters and atom
positions, and employs a powerful matching code~\cite{Ong2013} to categorize
materials within sensible tolerances. Each material is assigned a corresponding
unique material identifier that maps directly to web resources, can be used in
publications, and underpins the entire software/computing infrastructure.

A user's definition of a material might be broader or different than MP's;
the framework uses a hierarchy of material categories to allow the user to
attribute the contribution as is best suited to the dataset. At the top of the
hierarchy is the \textit{chemical space}
(\texttt{Fe\textsuperscript{*}O\textsuperscript{*}}), which includes many
\textit{compositions} (\texttt{FeO}, \texttt{Fe$_{\text{2}}$O$_{\text{3}}$},
\texttt{Fe$_{\text{3}}$O$_{\text{4}}$}) with the respective elements. Each
composition can have multiple \textit{structures}, which in turn correspond to
a unique \textit{MP material}.

Generally, a submission can be a contribution in any of the above categories.
For example, data received from researchers at LBNL's Advanced Light Source
typically refer to chemical spaces or compositions rather than materials. The
category for a given contribution is determined by the parser
(Sec.~\ref{sec_parsbld}) from the \texttt{MPFile}'s root-level section title
(Sec.~\ref{subsec_preproc_mpfile}).

\subsection{I/O File \texttt{MPFile}: Format \& Object}\label{subsec_preproc_mpfile}

In this section, we describe the \texttt{MPFile} format for the creation,
editing, organization, and storage of user-contributed data. The format
consolidates an experimental or theoretical contribution's tabular data
together with its meta-data. \texttt{MPFiles} are plain text files that are
designed to be easily read and edited by a person, yet also easily parsed and
processed by programs. An \texttt{MPFile} resembles the Investigation Study
and Assay Tabular format (ISATab)~\cite{isa-tools}, which is an interchange
format for experimental life sciences data with an associated tool suite and
REST API. Unlike ISATab, \texttt{MPFile} works as free-form data without
templates or ontologies and requires only a valid materials identifier.

\begin{SCfigure*} \centering \includegraphics[width=0.72\textwidth]{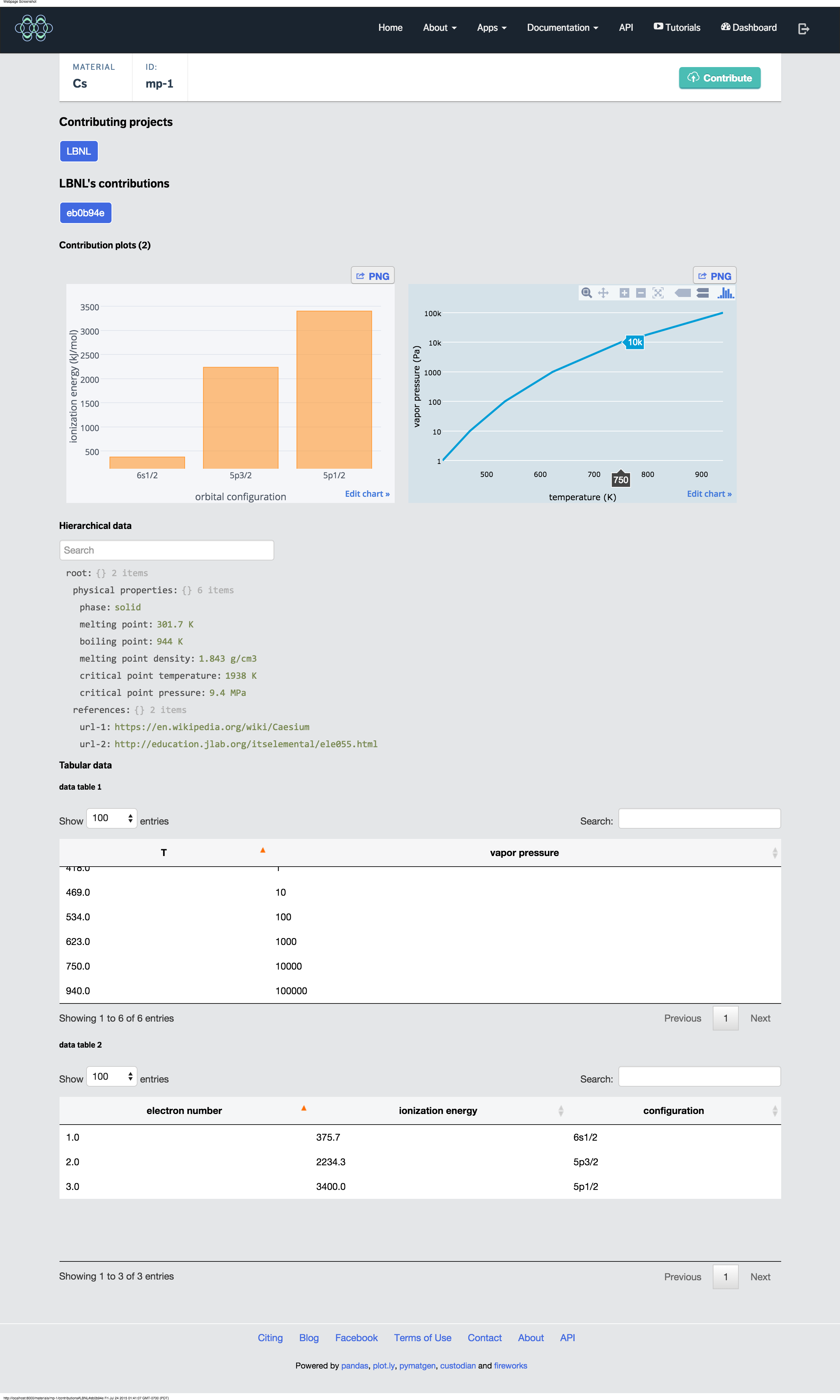}
\caption{Screenshot of \texttt{MPContribs}' front-end prototype showing the
  contribution with shortened identifier eb0b94e extracted from the \texttt{MPFile} contents
  of Example \ref{myautocounter} (first root-level section). The user can
  toggle different contributing projects and according contributions. For a
  selected contribution, interactive plots are shown along with its
  hierarchical and tabular data (top to bottom). The user's cursor is over the
  plot at right, triggering an interactive mode. Note the search field for
  hierarchical data that reduces the presented data to sub-dictionaries
  containing the search string. A real-time demonstration of the workflow
  enabled by \texttt{mgc} and the \texttt{MPContribs} infrastructure can be
  found at \url{https://youtu.be/lGwYLk2BlLA}.}
\label{fig_frontend}

\end{SCfigure*}

Our \texttt{MPFile} format could best be described as CSV sections wrapped and
extended by a minimal amount of meta-data necessary to customize the submission
for MP. The extensions allow one to link data to existing materials,
generate plots, and present hierarchical key-value pairs as well as free-form
text. Apart from being human-readable and -editable, the file reflects the
structural presentation on the front-end contribution page
(Fig.~\ref{fig_frontend}). These features, along with native CSV support, is a
major reason we chose to deviate from existing generic formats such as XML,
JSON, and YAML.

The \texttt{MPFile} format indicates header levels by repetitions of a special
character, a strategy borrowed from well-known markup languages such as
Markdown and Wikitext. Markdown, for instance, uses the `\texttt{\#}' character
for this purpose. We chose instead the `\texttt{>}' symbol to avoid conflicts
with the way comments are indicated in most CSV files in which the
`\textit{\#}' character is reserved for commenting. To avoid collision with the
mathematical `\texttt{>>}' sign and to improve visibility/readability, we use a
minimum repetition of three (`\texttt{>>>}'). Text following this section
delimiter is parsed as section name apart from comments appended on the same
line.


Example \ref{myautocounter} shows a sample \texttt{MPFile} for the contribution
of data on basic physical properties for caesium and palladium. The MP internal
identifiers for caesium and palladium are \texttt{MP-1} and \texttt{MP-2},
respectively, which are used as section names to associate the contributed data
with the corresponding entry in the core dataset. The same form of root-level
section headers applies to composition identifiers.

The two line types are section headers (lines 1, 2, 9, etc.) and section
contents. The section headers may be a reserved word (described below) or an
arbitrary property title. The section contents can be either key/value pairs or
a table of comma-separated values. The key/value pairs are used to organize
meta-data or annotated numbers in free form using section names that best
describe the contained data (lines 3--8, 10--11, 14--17). Possibly multiple tables of data in CSV
format, including a header row with column names, are embedded in level-1 sections
(lines 19--25, 27--30, 33--39). Note that there is no requirement
for \texttt{MPFile} templates that enforce common key or column names.  This
provides maximum flexibility to the user while still enabling support for the
hierarchy of Structure, Composition, and Chemical Space
(Sec.~\ref{subsec_preproc_matdef}).  These restrictions can always be added
later in response to a genuine need and independent agreement from within the
user community.
We also note that the infrastructure is geared towards kilobytes and not
megabytes of data per submission. Large raw data can instead be linked to in
the meta-data.

\begin{pabox}[label={myautocounter}]{Sample \texttt{MPFile} content for two
materials. Line numbers and bold section headers were added for presentation
purposes. The web page for the \texttt{MP-1} section is shown in
Figure~\ref{fig_frontend}.}
\tcbfontsize{0.73}%
\setlength\linenumbersep{5pt}%
\internallinenumbers%
\textbf{\texttt{>>>} MP-1 \# caesium}\\
\textbf{\texttt{>>>>} physical properties}\\
phase: solid\\
melting point: 301.7 K\\
boiling point: 944 K\\
melting point density: 1.843 g/cm3\\
critical point temperature: 1938 K\\
critical point pressure: 9.4 MPa\\
\textbf{\texttt{>>>>} references} \# list of url, bibtex, or doi\\
url-1: ``https://en.wikipedia.org/wiki/Caesium''\\
url-2: ``http://education.jlab.org/itselemental/ele055.html''\\
\textbf{\texttt{>>>>} plots}\\
\textbf{\texttt{>>>>>} default data table 2} \# overwrite graph properties\\
x: configuration\\
y: ionization energy\\
kind: bar\\
table: table 2\\
\textbf{\texttt{>>>>} table 1} \# can be named freely\\
T, vapor pressure\\
418,1\\
469,10\\
534,100\\
623,1000\\
750,10000\\
940,100000\\
\textbf{\texttt{>>>>} table 2}\\
electron number, ionization energy, configuration\\
1,375.7,6s1/2\\
2,2234.3,5p3/2\\
3,3400,5p1/2\\
\\
\textbf{\texttt{>>>} MP-2 \# palladium} (bare data section)\\
temperature (K), vapor pressure (Pa)\\
1721,1\\
1897,10\\
2117,100\\
2395,1000\\
2753,10000\\
3234,100000
\end{pabox}

The reserved section titles \textit{general}, \textit{plots}, and
\textit{references} indicate special significance and processing:

A \textit{general} section (omitted in the example) is different from normal
sections with key/value pairs only in two regards. In the future, MP
might require certain unique key names in this section: a mandatory description
field, for instance, to provide an informative overview of the contribution on
the front-end. More importantly, the \textit{general} section is employed to
support an \texttt{MPFile} mode in which common meta-data is shared amongst
contributions on multiple materials and/or datasets.
Section~\ref{subsec_parsbld_recparse} describes the internal implementation of
this feature during the \texttt{MPFile} parser phase.

By default, a two-dimensional graph is generated for each CSV table with the
first column as the abscissae and each remaining column as corresponding
ordinates for as many traces.  The optional \textit{plots} section can be used
to request additional graphs or overwrite properties of the default ones based
on the sections containing CSV tables. The plot axes are selected or changed by
referring to the header of the table indicated in the \textit{table} sub-key
(lines 13--17).  The keys chosen for this section (\textit{x}, \textit{y},
\textit{kind}, etc.) are the same as those used by the Pandas Python library
(through cufflinks~\cite{cufflinks}).  Details on plot generation using Plotly
are given in Section~\ref{subsec_parsbld_build}.

The \textit{references} section (lines 9--11) allows for the attribution of
credit and for proper provenance of the submitted datasets. It only accepts a
list of key/value pairs with the keys \texttt{url}, \texttt{doi}, or
\texttt{bibtex}. The BibTeX string can either be obtained from the DOI or
entered manually, and is subsequently resolved dynamically for prominent
display on the front-end.

\subsection{Collaborative Extension: \texttt{MPContribsUsers}}\label{subsec_preproc_users}

The \texttt{MPFile} format can be parsed and manipulated by Python
functions/classes in the open-source \texttt{MPContribs} library. This allows
users to construct \texttt{MPFiles} manually or through their own custom
scripts. To coordinate between users and reduce duplication of work from
writing code that deals with similar or identical formats in different domains
of materials science, we created \texttt{MPContribsUsers}~\cite{mpcontribs}, a
codebase hosted on Github containing reusable routines for both data generation
and data analysis of \texttt{MPFiles} (c.f. Fig.~\ref{fig_flow_chart}, left and
right of REST API ``wall'', respectively).

To engage the developer community, quality code practices are important, so for
both core and extension repositories we use Github for source code control and
issue tracking, and include extensive unit testing and documentation in the
code, and in the project using version-controlled Markdown text.



The routines for data generation transform output files from common simulation
software or from experimental instruments into the \texttt{MPFile} format.
For instance, \texttt{MPContribsUsers} can assimilate data from the
\textit{VASP} software~\cite{vasp} by extending tools built in
\textit{pymatgen}. Other routines aid in determining the appropriate materials
identifier for a contribution.

The collected code for data analyses performs \textit{joint analyses} of MP
core data and \texttt{MPFile} contributions. For instance, submitted
experimental XAS spectra could be studied with respect to
FEFF~\cite{feff} calculations to relate the measured compositions to possible
crystal structures. Similarly, XMCD signals resulting from the XAS spectra can
be overlaid with the corresponding theoretical phase diagrams calculated by MP,
providing a unified view of experiment and theory. The workflow of this
\textit{Unified Theoretical and Experimental x-ray Spectroscopy Application} in
addition to \texttt{MPContribs}' role in the \textit{Nanoporous Materials
  Explorer} are discussed in~\cite{mpcontribs_gce_short}. Note that such
dedicated analysis applications within \texttt{MPContribsUsers} also drive the
feedback loop of derived data back into the core and contribution collections.
The collections can subsequently underpin project-specific web applications
with capabilities beyond the generic display.

%

\subsection{Internal Contribution Processing}\label{sec_parsbld}

The two main phases of the internal contribution processing are the parsing of
the submitted \texttt{MPFile} (Sec.~\ref{subsec_parsbld_recparse}), and the
(re-)aggregation of single contributions (Sec.~\ref{subsec_parsbld_build}) to
build derived collections and generate interactive Plotly graphs.

\subsubsection{\texttt{MPFile} Parser}\label{subsec_parsbld_recparse}
The purpose of the \texttt{MPFile} parser is to (1) handle both the
multiple-material and multiple-dataset contribution modes, (2) split the
contribution data into ``atomic'' pieces defined by materials identifier, and
(3) store each in a dedicated database with a unique
contribution identifier. The underlying algorithm proceeds by
recursively unpacking the \texttt{MPFile}'s section contents with increasing
section level and hence allows for arbitrarily deep section nesting.
For the multiple-dataset contribution mode (Sec.~\ref{subsec_preproc_mpfile}),
the first root-level section denotes the main \textit{general} section
which is applied to all subsequent root-level sections. This effectively merges
the two contribution modes and allows for meta-data to be shared amongst
multiple datasets with local \textit{general} subsections taking precedence.
All section contents are imported into the internal recursive updating
dictionary and data validation/cleaning handled by taking advantage of the
tools provided by the Pandas library~\cite{pandas}.
Pandas supports the import of data from many sources which makes it a suitable
basis for the future implementation of additional features.

%

\subsubsection{(Re-)Aggregation and Plot Generation}\label{subsec_parsbld_build}

Derived collections are subsequently built from the canonical contributions
collection by (re-)aggregating the single contributions for a specific
materials identifier from many contributors. The resulting materials and
composition collections serve as direct query access points for the front-end.
In addition to ensuring that contributions are connected to the appropriate
materials, this \textit{builder} phase creates interactive plots from the user
data. The graphs and their options, requested by the user in the
\texttt{MPFile}'s \textit{plots} subsection, are converted into an interactive
graph through a HTTP request to Plotly's REST API~\cite{plotly}. The returned
unique URL is saved in the derived collections and used to embed the graph on
MP's front-end (c.f. Fig.~\ref{fig_frontend}). This allows direct access to the
graph online to edit it to the contributor's liking on Plotly's workspace. The
framework's integration with Pandas and Plotly during this builder phase is
smart enough not to overwrite the customization of existing graphs when
contributions are updated.

\section{Submission Display and Dissemination}\label{sec_webacc}

Submissions are accessible to users through a command-line client, an
interactive web UI, and a RESTful API. This section describes each of these
interfaces in more detail.

\subsection{Command-line Interface}\label{subsec_mgc}

The command-line client interface, called \texttt{mgc} (for \textit{Materials
Genome Contribution}), uses the MP's REST API
(Sec.~\ref{subsec_webacc_rest}) to submit user-contributed data to MP
for immediate visualization and dissemination to its users. The program accepts
a variety of positional arguments to facilitate the contributor's submission
management.
For instance, the \texttt{submit} sub-command enables the submission of a
physical \texttt{MPFile}. The initial submission embeds the generated
contribution identifiers into the \texttt{MPFile} to facilitate book-keeping
and to automatically trigger data updates upon resubmission.

\subsection{Web UI}\label{subsec_webacc_frontend}

Our front-end interface provides access and display of contributed data in the
context of the main MP web pages. Given the large number of users of the MP
website, it is important that contributors can look at and optimize the final
form themselves before public release of the contributed data. Thus, we provide
an incubation period during which visibility of a given contribution is
restricted to its contributor and invited collaborators. The primary task
during this period is for contributors to ensure that the plots presented
adhere to their standards for proper visual encoding of the data. Our plans
for the system to enable quality control other than that pertaining to
formatting are discussed in Sec. \ref{sec_out}.

Furthermore, while an incubation period provides a temporary means of access
control, we intend data submitted via \texttt{MPContribs} to ultimately be openly
accessible. Data contributions for activities where the quality of access
controls are critical are not well-suited for \texttt{MPContribs}; however, a scheme for
``sandboxed'' data access control has been implemented with success by MP for
such situations~\cite{qu_eg_2015}.

We seek to accommodate a variety of implicit data schemas from contributors.
Figure~\ref{fig_frontend} depicts a minimal starting point for flexible display
of user contributions that consist of arbitrary data in both hierarchical and
tabular forms, along with a collection of plots. For the material depicted, a
user may toggle among the contributing projects and their associated
contributions. This allows a user/institution to quickly browse its own
contributions for a material and also directly link to them, ultimately via
assigned DOIs.

The interface encourages exploration. A tree of hierarchically structured data
expands incrementally as characters are typed into its search field, allowing
one to quickly search what may be a formidable hierarchy of key/value pairs.
The tabular data display features an incremental-search field as well; the
table is both paginated and scrollable.
A contributing project's plots each link to the Plotly~\cite{plotly} workspace,
so that users can jump directly into editing the graphs.

The built-in extensibility and portability of the plots served with each
contribution (through Plotly) help users present explorable explanations of
their contributions and retrieve publication-ready static assets. Optional
widgets directly customize plot interactivity on the MP site, and plots can be
transparently exported to a variety of file formats through simply adding an
appropriate file extension to their URL. This portability also ensures that
bandwidth-efficient static images of plots are served for small-screened mobile
devices.


\subsection{RESTful API}\label{subsec_webacc_rest}

The implementation of \texttt{MPContribs}' REST API makes use of MP's
existing code base to expose
the I/O capabilities provided by the \texttt{MPFile} object
(Sec.~\ref{subsec_preproc_mpfile}) and the database interface class.
For example, we have implemented a general query function
that can search based on \texttt{MPFile} section names and attributes. The
resulting HTTP endpoints can be used via any language's HTTP library, but for
Python dedicated convenience methods have been implemented as part of
pymatgen's \texttt{MPRester} class.

Once we have a substantial collection of user
contributions on which to practice inference of implicit schemas, or
alternatively can allow users to explicitly assign known type classes to
elements of their data, we can develop tools to enable further analyses of
contribution data integrated with MP core data. Such tools, however, need to be
developed with care to avoid limiting the framework's flexibility of data
ingestion.

Note that the effort to provide a convenient REST API for \texttt{MPContribs}
accelerates the development progress and user acceptance of the framework. It
also allows the framework to be independently developed after it is released.

\section{Summary}\label{sec_sum}

As the rate of data generation in the materials science community continues to
grow, the need for efficient and collaborative vehicles for data sharing
becomes increasingly evident. The internet's history has shown that such
continued growth is only possible by directly involving the community in data
submission and vetting. In this paper, we introduced the Materials Project
contributions framework, a software platform that allows users to share data
with over 10,000 users.

A major design challenge is to link data pertaining to the same ``material'', as
the identity of a real material can often neither be fully specified nor even
rigorously defined. \texttt{MPContribs} utilizes the hierarchical nature of a
material specification to intrinsically cover various contribution scopes: from
computational data with full structural information to experimental results
with only partial material descriptions.

The framework seamlessly integrates data composed of key/value pairs, data
tables, nested data structures, and interactive plots. These features provide
considerable flexibility in how users can contribute and present their data to
the community. The framework implementation merges existing technologies with
the custom but generic and human-readable \texttt{MPFile} format. Its parser
supports two common submission modes and decomposes a contribution into
suitable ``atomic'' MongoDB documents. Post-processing steps intelligently
combine contributions from many users such that contributions belonging
together can easily be rendered and served instantly to the public by the
contribution-agnostic front-end.

Integration of the data handling with an existing collaborative graphing
platform puts organization, maintainance, and formatting under users' control.
A dynamic programmable REST API not only provides mechanisms for book-keeping
and retrieval of submitted entries but also aggregates submissions as needed
for integrated analyses that use metrics across contributed and core datasets.
Collaboration on raw data processing and integrated analyses with the developer
team is enabled by the separate \texttt{MPContribsUsers} area which serves as a
growing repository of well-maintained user codes.

Our \texttt{MPContribs} toolset allows the materials researchers to present
curated datasets to the broader community, and to tightly integrate them with
those already available on MP. For developers, the presented solution provides a path
from prototype to production with managable implementation efforts. Our effort
is distinguished from others in that its goal is not to preserve and store
large data files in their native format, but rather to provide a curated view
of structured and processed data that is suitable for direct analysis and query
and is integrated within an existing materials database.

\section{Outlook}\label{sec_out}

We expect future work to be largely guided by user requirements determined
through iteration and testing. However, several challenges are already apparent
and our strategy to handle them will evolve as the number of contributors
grows.

\textit{A)} Defining universal lexicons for the description of materials and
their properties across a broad user base is needed to unify the common
elements of the datasets so that they can be queried and analyzed together.

\textit{B)} Providing users with the ability to validate their data through
continuous integration would greatly improve the robustness and reliability of
user-submitted datasets. Currently, we closely collaborate with and limit
contributions to trusted domain experts to produce curated, high-quality
datasets. In this first stage, we aim to provide tools for efficient manual
data inspection. For instance, a separate ``staging'' area for data
contributions is already in place~\cite{mpcontribs_gce_short} and allows for
human verification before release on the live web site. Further, our
integration of the plotting tool Plotly quickly reveals incorrect data such as
outliers. In the next stage, we will move from human to automatic
self-verification of the data: we plan to release a data access API along with
instructions on how users can employ the API to develop unit testing and
continuous integration frameworks themselves. If problems of data quality
persist, we will proceed to a third stage of development that allows users to
rate each other's dataset quality which enables mechanisms like flagging,
removal or ``black-listing'' of consistently low-quality datasets/users.
  
\textit{C)} As the project scales to storing more data in a centralized manner,
the issue of preventing data loss and scaling database performance/storage will
become increasingly challenging.

However, despite these challenges, and although our
framework is still in its early stages, feedback from several institutions has
been positive and our project already has a wait-list of several research
groups awaiting further developments. Therefore, we are confident of the need
for a community-wide framework for sharing materials datasets and are
optimistic about the potential of data sharing to accelerate scientific
discovery and innovation.

\section*{Acknowledgment}

This work was intellectually led by the Department of Energy's Basic Energy
Sciences program - the Materials Project - under Grant No. EDCBEE and supported
by Center for Next Generation Materials by Design, an Energy Frontier Research
Center funded by DOE, Office of Science, BES. Work at Lawrence Berkeley
National Laboratory was supported by the Office of Science of the U.S.
Department of Energy under Contract No. DEAC02-05CH11231. We thank the National
Energy Research Scientific Computing Center for providing invaluable computing
resources. Finally, we would like to thank all the users of the Materials
Project for their support and feedback in improving the project.

\IEEEtriggeratref{50}
\bibliographystyle{IEEEtran}
\bibliography{IEEEabrv,eScience15}

\begin{thebibliography}{10}
\providecommand{\url}[1]{#1}
\csname url@samestyle\endcsname
\providecommand{\newblock}{\relax}
\providecommand{\bibinfo}[2]{#2}
\providecommand{\BIBentrySTDinterwordspacing}{\spaceskip=0pt\relax}
\providecommand{\BIBentryALTinterwordstretchfactor}{4}
\providecommand{\BIBentryALTinterwordspacing}{\spaceskip=\fontdimen2\font plus
\BIBentryALTinterwordstretchfactor\fontdimen3\font minus
  \fontdimen4\font\relax}
\providecommand{\BIBforeignlanguage}[2]{{%
\expandafter\ifx\csname l@#1\endcsname\relax
\typeout{** WARNING: IEEEtran.bst: No hyphenation pattern has been}%
\typeout{** loaded for the language `#1'. Using the pattern for}%
\typeout{** the default language instead.}%
\else
\language=\csname l@#1\endcsname
\fi
#2}}
\providecommand{\BIBdecl}{\relax}
\BIBdecl

\bibitem{WHMGI2014}
\BIBentryALTinterwordspacing
{Materials Genome Initiative}. [Online]. Available: \url{www.whitehouse.gov}
\BIBentrySTDinterwordspacing

\bibitem{Jain2013}
A.~Jain \emph{et~al.}, ``{The Materials Project: A materials genome approach to
  accelerating materials innovation},'' \emph{APL Mat. 1}, p. 011002, 2013.

\bibitem{Berman2000}
\BIBentryALTinterwordspacing
H.~M. Berman \emph{et~al.}, ``{The Protein Data Bank},'' \emph{Nucl. Acids
  Res.}, vol.~28, pp. 235--242, 2000. [Online]. Available:
  \url{www.rcsb.org/pdb}
\BIBentrySTDinterwordspacing

\bibitem{Belsky2002}
A.~Belsky \emph{et~al.}, ``{New ICSD Developments: Accessibility in Support of
  Materials Research and Design},'' \emph{Acta Cryst. B58}, pp. 364--369, 2002.

\bibitem{pauling}
\BIBentryALTinterwordspacing
{Pauling File}. [Online]. Available: \url{http://paulingfile.com}
\BIBentrySTDinterwordspacing

\bibitem{aflowlib}
\BIBentryALTinterwordspacing
{AFLOW for Materials Discovery}. [Online]. Available: \url{www.aflowlib.org}
\BIBentrySTDinterwordspacing

\bibitem{cccbdb}
\BIBentryALTinterwordspacing
{NIST CCCBDB}. [Online]. Available: \url{http://cccbdb.nist.gov}
\BIBentrySTDinterwordspacing

\bibitem{comp-es}
\BIBentryALTinterwordspacing
{CompES Database}. [Online]. Available: \url{http://caldb.nims.go.jp}
\BIBentrySTDinterwordspacing

\bibitem{ortiz2009data}
C.~Ortiz \emph{et~al.}, ``Data mining and accelerated electronic structure
  theory as a tool in the search for new functional materials,'' \emph{Comp.
  Mat. Sci.}, vol.~44, no.~4, pp. 1042--1049, 2009.

\bibitem{cepdb}
\BIBentryALTinterwordspacing
{Clean Energy DB}. [Online]. Available: \url{http://cepdb.molecularspace.org}
\BIBentrySTDinterwordspacing

\bibitem{oqmd}
\BIBentryALTinterwordspacing
{Open Quantum Materials Database}. [Online]. Available: \url{http://oqmd.org}
\BIBentrySTDinterwordspacing

\bibitem{nomad}
\BIBentryALTinterwordspacing
{NoMaD Repository}. [Online]. Available: \url{http://nomad-repository.eu}
\BIBentrySTDinterwordspacing

\bibitem{moreno2015managing}
\BIBentryALTinterwordspacing
M.~Alvarez-Moreno \emph{et~al.}, ``{Managing the Computational Chemistry Big
  Data Problem: The ioChem-BD Platform},'' \emph{J. Chem. Inf. Model.},
  vol.~55, pp. 95--103, 2015. [Online]. Available: \url{http://iochem-bd.org}
\BIBentrySTDinterwordspacing

\bibitem{mdf}
\BIBentryALTinterwordspacing
{Mater. Data Facility}. [Online]. Available: \url{www.nationaldataservice.org}
\BIBentrySTDinterwordspacing

\bibitem{iplant}
\BIBentryALTinterwordspacing
{{iPlant} Collaborative}. [Online]. Available:
  \url{www.iplantcollaborative.org}
\BIBentrySTDinterwordspacing

\bibitem{kbase}
\BIBentryALTinterwordspacing
{Systems Biology Knowledgebase}. [Online]. Available: \url{https://kbase.us}
\BIBentrySTDinterwordspacing

\bibitem{perez2007ipython}
F.~P\'erez and B.~E. Granger, ``{{IP}ython: a System for Interactive Scientific
  Computing},'' \emph{Comp. Sci. Eng.}, vol.~9, no.~3, pp. 21--29, 2007.

\bibitem{chemspider}
\BIBentryALTinterwordspacing
{ChemSpider}. [Online]. Available: \url{www.chemspider.com}
\BIBentrySTDinterwordspacing

\bibitem{Heller2013}
S.~Heller \emph{et~al.}, ``{InChI - the worldwide Chemical Structure Identifier
  Standard},'' \emph{J Chem. Inf. Model.}, vol.~5, p.~7, 2013.

\bibitem{mpcontribs}
\BIBentryALTinterwordspacing
P.~Huck. {Materials Project User Contribution Framework}. [Online]. Available:
  \url{{https://github.com/materialsproject/MPContribs{Users}}}
\BIBentrySTDinterwordspacing

\bibitem{Ong2013}
S.~P. Ong \emph{et~al.}, ``{PyMatGen: A robust, open-source Python Library for
  Materials Analysis},'' \emph{Comp. Mat. Sci.}, vol.~68, pp. 314--319, 2013.

\bibitem{isa-tools}
\BIBentryALTinterwordspacing
{ISA MetaData Tracking}. [Online]. Available: \url{http://www.isa-tools.org}
\BIBentrySTDinterwordspacing

\bibitem{cufflinks}
\BIBentryALTinterwordspacing
J.~Santos. {Productivity Tools for Plotly + Pandas}. [Online]. Available:
  \url{https://github.com/santosjorge/cufflinks}
\BIBentrySTDinterwordspacing

\bibitem{vasp}
\BIBentryALTinterwordspacing
{VASP}. [Online]. Available: \url{http://rcc.its.psu.edu/resources/software}
\BIBentrySTDinterwordspacing

\bibitem{feff}
J.~J. Rehr \emph{et~al.}, ``{Parameter-free calculations of X-ray spectra with
  FEFF9},'' \emph{Phys. Chem. Chem. Phys.}, vol.~12, pp. 5503--5513, 2010.

\bibitem{mpcontribs_gce_short}
P.~Huck \emph{et~al.}, ``{User Applications Driven by the Community
  Contribution Framework MPContribs in the Materials Project},''
  \emph{Concurrency Computat.: Pract. Exper. (submitted)}, vol. {10th GCE
  Workshop}, 2015.

\bibitem{pandas}
\BIBentryALTinterwordspacing
{Pandas Python Library}. [Online]. Available: \url{http://pandas.pydata.org}
\BIBentrySTDinterwordspacing

\bibitem{plotly}
\BIBentryALTinterwordspacing
{Collab. Graphing Platform Plot.ly}. [Online]. Available: \url{https://plot.ly}
\BIBentrySTDinterwordspacing

\bibitem{qu_eg_2015}
X.~Qu \emph{et~al.}, ``{The Electrolyte Genome project: A big data approach in
  battery materials discovery},'' \emph{Comp. Mat. Sci.}, vol. 103, pp. 56--67,
  2015.

\end{thebibliography}

\end{document}